\documentclass[lettersize,journal]{IEEEtran}
\usepackage[font={small}]{caption}
\usepackage[table]{xcolor}
\usepackage{amsmath,graphicx}
\usepackage{algorithm}
\usepackage{algorithmicx}
\usepackage{algpseudocode} 
\usepackage{array}
\usepackage{subfig}
\usepackage{textcomp}
\usepackage{stfloats}
\usepackage{siunitx}
\usepackage{url}
\usepackage{siunitx}
\usepackage{verbatim}
\usepackage{graphicx}

\usepackage{hyperref}
\hypersetup{
    colorlinks=true,
    linkcolor=blue,
    filecolor=blue,      
    urlcolor=blue,
    citecolor=blue,
}

\usepackage{mathrsfs}
\usepackage{bm}
\usepackage{amssymb}
\usepackage{multirow}
\usepackage{booktabs}
\usepackage{threeparttable} 
\usepackage[numbers,sort&compress]{natbib}
\begin{document}

\title{Zero-Shot Semantic Communication with Multimodal Foundation Models
\thanks{J. Hu, W. Xu, W. Zhang, F. Wang, H. Gao are with the Beijing University of Posts and Telecommunications, Beijing 100876, China.  (Email: fengyu.wang@bupt.edu.cn). H. Wu and D. G{\"u}nd{\"u}z are with the Department of Electrical and Electronic Engineering, Imperial College London, London SW7 2AZ, U.K. This work was carried out when J. Hu and W. Zhang were visiting the Information Processing Laboratory at Imperial College London.}

}


\author{Jiangjing Hu, Haotian Wu, Wenjing  Zhang, Fengyu Wang, Wenjun Xu,  \IEEEmembership{Senior Member, IEEE},\\ Hui Gao, \IEEEmembership{Senior Member, IEEE}, Deniz G{\"u}nd{\"u}z, \IEEEmembership{Fellow, IEEE}}


\maketitle

\begin{abstract}
\textcolor{black}{Most existing semantic communication (SemCom) systems use deep joint source-channel coding (DeepJSCC) to encode task-specific semantics in a goal-oriented manner. However, their reliance on predefined tasks and datasets significantly limits their flexibility and generalizability in practical deployments. Multi-modal foundation models provide a promising solution by generating universal semantic tokens. Inspired by this, we introduce SemCLIP, a zero-shot SemCom framework leveraging the contrastive language-image pre-training (CLIP) model. By transmitting CLIP-generated image tokens instead of raw images, SemCLIP enables efficient SemCom under low bandwidth and challenging channel conditions, facilitating diverse downstream tasks and zero-shot applications. Specifically, we propose a DeepJSCC scheme for efficient CLIP token encoding. To mitigate potential degradation caused by compression and channel noise, a multi-modal transmission-aware prompt learning mechanism is designed at the receiver, which adapts prompts based on transmission quality, enhancing system robustness and channel adaptability. Simulation results demonstrate that SemCLIP outperforms the baselines, achieving a $41\%$ improvement in zero-shot performance at low signal-to-noise ratios. Meanwhile, SemCLIP reduces bandwidth usage by more than $50$-fold compared to alternative image transmission methods, demonstrating the potential of foundation models towards a generalized, task-agnostic SemCom solution.}


\end{abstract}


\section{Introduction}

\IEEEPARstart{I}{n} the era of the sixth-generation~(6G) mobile networks, communications among humans, machines, and intelligent agents enable diverse intelligent tasks, increasing the demand for efficient information transmission. Recent results on semantic/goal-oriented communications have highlighted the importance of extracting and transmitting only the most relevant information for the task that the receiver wants to carry out~\cite{6g}. In current separation-based communication systems, the objective is only relevant for data compression, while channel transmission is designed independent of the data distribution or task objective~\cite{6g}. However, in practical finite block-length systems, the separation-based approach is suboptimal, and a joint source-channel coding~(JSCC) design is needed~\cite{wu2025deep}.

Most of the current semantic communication (SemCom) systems utilize deep joint source-channel coding~(DeepJSCC) to encode task-related semantics for efficient transmission over noisy channels in an end-to-end~(E2E) fashion~\cite{jscc1-retrival,ib,lo2023collaborative}. Particularly, a DeepJSCC scheme is proposed in~\cite{jscc1-retrival} to communicate retrieval-oriented semantic features. DeepJSCC parameters are optimized in~\cite{ib} guided by the information bottleneck principle. Meanwhile, a collaborative DeepJSCC is designed in~\cite{lo2023collaborative} to perform edge inference. Multi-task transmission is also explored in~\cite{multi-task,multi-task2,multi-task3}, including multi-task cooperative DeepJSCC~\cite{multi-task}, graph attention-based transmission~\cite{multi-task2}, and multi-modal scenarios~\cite{multi-task3}. 
Nevertheless, all of the above systems are trained for a specific and predetermined list of tasks on a particular dataset even if only one task is chosen randomly at the time of inference. Therefore, retraining is inevitable for new scenarios, limiting their flexibility and hindering the development of a generalizable SemCom system.

To enhance the generalizability of SemCom, most of the existing works rely on transfer learning to transfer semantic information across different scenarios and tasks. In~\cite{transfer}, a semantic encoder is pre-trained for image classification, and subsequently fine-tuned using a few labels for image object detection. To facilitate a task-agnostic system, an auxiliary data adaptation network is introduced in~\cite{task-unkown}, which utilizes domain adaptation techniques for knowledge transfer. Alternatively, \cite{zero-shot} explores a semantic knowledge base with a multi-level feature extractor, formulating dynamic feature transmission as a binary knapsack problem. However, all the above approaches still require additional training based on some prior knowledge or structural modifications to adapt to new scenarios, increasing system complexity. 


To further improve the universality and robustness of semantic representations, foundation models can be integrated into SemComs. For example, stable diffusion model~\cite{diffusion} enables high-fidelity generative modeling by iteratively removing noise from latent representations, while generative pre-trained transformer~(GPT)~\cite{achiam2023gpt} allows contextual reasoning over combined image and text inputs. Meanwhile, multi-modal foundation models, such as the contrastive language-image pre-training (CLIP) model~\cite{clip}, is trained on extensive image-text pairs to learn general visual concepts from natural language supervision. On this basis, generative foundation models and large language models can be introduced to improve semantic coding effectiveness~\cite{foundationsurvey2}. In~\cite{qiao}, the authors enhance the transmission efficiency by decomposing the signal into multiple modalities, which are transmitted separately, and combined at the receiver using a pre-trained diffusion model. However, these works ignore channel noise-aware design, degrading transmission efficiency.

\color{black}
In this correspondence, we consider a zero-shot SemCom scenario, in which the transmitter has no prior knowledge regarding the task to be carried out at the receiver; and hence, cannot implement any fine-tuning or adaptation for the task at hand. One potential solution in this scenario would be to transmit the input sample in a lossless fashion, allowing the receiver to carry out any desired tasks. However, this would require significant bandwidth and result in the transmission of features that are irrelevant to most tasks. \textcolor{black}{To address these challenges and develop a zero-shot SemCom framework, we propose SemCLIP, a SemCom scheme based on the CLIP model. Leveraging the generalizability of CLIP, SemCLIP communicates CLIP-based image tokens via a signal-to-noise ratio~(SNR)-adaptive DeepJSCC scheme, enabling zero-shot applications in a task-agnostic manner.}
Our key contributions are as follows:
\begin{itemize}
\item{
We propose a CLIP-based SNR-adaptive JSCC scheme to enable the efficient transmission of image tokens in a task-agnostic fashion for zero-shot remote tasks. The proposed method enhances transmission robustness against channel noise while preserving generalizability. Notably, it serves as a unified approach and can be extended to token transmissions for other foundation models.
}
\item{To mitigate the impact of channel noise and improve task performance, a transmission-aware prompt learning~(TAPL) mechanism is proposed, where text prompts are adjusted adaptively according to the JSCC-decoded features. }
\item{Simulation results show that the proposed method outperforms baselines with up to $41\%$ improvement in zero-shot performance and $50$-fold bandwidth reduction in transmission efficiency across different datasets, highlighting the potential of foundation models towards a generalized, task-agnostic SemCom solution.}
\end{itemize}

\color{black}
\vspace{-1em}
\section{{System Model}}
\textcolor{black}{We consider a zero-shot remote task scenario where images are transmitted in a task-agnostic manner. 
}
At the transmitter, a batch of $B$ images $\bm{x}\in\mathbb{R}^{B \times H \times W\times C}$ are encoded into channel symbols $\bm{z}\in\mathbb{C}^{B \times L}$, where $H$, $W$, and $C$ are the height, width, and color channels, respectively, and $L$ represent the number of channel uses per image. We impose an average power control on the channel input as $\frac{||{\bm{z}}||^2}{BL}\leq P$, where $P$ is the average power budget per symbol, which is set to $1$ without loss of generality. The encoding process is denoted as
$\bm{z}=\mathcal{E}(\bm{x})$.   

Subsequently, $\bm{z}\in\mathbb{C}^{B \times L}$ is transmitted over an additive white Gaussian noise~(AWGN) channel as:
\begin{equation}
{\bm{\hat{z}}}={\bm{z}}+\bm{n}, 
\end{equation}
where $\bm{n}\in \mathbb{C}^{B \times L}$ is the white Gaussian noise, with each element ${n}_i\sim\mathcal{C}\mathcal{N}(0,\sigma^2)$. The channel SNR is defined as $\mu=10\log_{10} \frac{P}{\sigma^2}$ dB, available at both the transmitter and receiver. 

At the receiver, channel output ${\bm{\hat{z}}}\in \mathbb{C}^{B \times L}$ is used for various tasks. Unlike other goal-oriented SemCom systems, images here are transmitted in a task-agnostic manner, preventing the system from fine-tuning or optimizing for specific objectives. There are two alternative approaches: 

(1) Image reconstruction~(IR)-based methods: Receiver reconstructs the input images as $\bm{\hat{x}}$ with the best possible quality, and performs downstream tasks to obtain the final result $\bm{\hat{p}}$, formulated as:
$\bm{\hat{x}}=\mathcal{D}_1(\bm{\hat{z}})$, $\bm{\hat{p}}=\mathcal{D}_2(\bm{\hat{x}})$.

(2) Feature transmission~(FT)-based methods: Receiver decodes unified semantic features $\bm{\hat{s}}$ with sufficient information for various tasks, without requiring image reconstruction:
$\bm{\hat{s}}=\mathcal{D}_1(\bm{\hat{z}}),  \bm{\hat{p}}=\mathcal{D}_2(\bm{\hat{s}})$.

For both methods, $\mathcal{D}_1(\cdot)$ represents the reconstruction/decoding process, and $\mathcal{D}_2(\cdot)$ performs the downstream task. The optimization objective is $\arg \min_{\mathcal{E},\mathcal{D}_1}\mathbb{E}\big[d(\cdot)\big]$, where $d(\cdot)$ measures reconstruction distortion—$d(\bm{x},\bm{\hat{x}})$ for IR and $d(\bm{s},\bm{\hat{s}})$ for FT. The expectation is computed across different images and channel distributions. The IR approach preserves maximum image information for downstream tasks, but incurs high communication costs. In contrast, FT prioritizes the universality of extracted semantics with lower bandwidth costs but faces challenges in ensuring semantic robustness across tasks and noisy transmissions.




\vspace{-1em}
\section{Design of SemCLIP}\label{}
We propose SemCLIP, a FT-based SemCom framework integrating foundation models (see Fig. \ref{SemCLIP framework}). Leveraging CLIP as a unified semantic encoder with a JSCC transmission scheme, \textcolor{black}{we validate SemCLIP’s generalizability through zero-shot classification and zero-shot cross-modal retrieval.} \textcolor{black}{Here, zero-shot cross-modal retrieval is performed at the receiver: a user-provided text query is used to identify the most semantically relevant image feature vector among the received ones. The index of the selected vector is then sent back to the transmitter, potentially triggering further actions (e.g. transmitting the full-resolution image). Our focus is on the retrieval stage, which enables efficient communication by avoiding transmission of the entire image dataset.}



\vspace{-1.3em}
\subsection{Workflow}
At the transmitter, image tokens $\bm{s}\in\mathbb{R}^{B \times N}$ are extracted from $B$ query images by a CLIP-based semantic encoder, where $N$ is the feature dimension. An SNR-adaptive JSCC-encoder then maps $\bm{s}$ to $\bm{z}$, which is transmitted over the channel. 
At the receiver, the SNR-adaptive JSCC-decoder reconstructs image tokens $\bm{\hat{s}}\in\mathbb{R}^{B \times N}$ from the received noisy vector ${\bm{\hat{z}}}$. \textcolor{black}{Subsequently, $\bm{\hat{s}}$ is fed into a TAPL-based task performer for multi-modal zero-shot classification or retrieval, whose details are provided in Subsection~\ref{tapl}}. 

Specifically, the prediction probability is obtained by computing the semantic similarity between $\bm{\hat{s}}$ and a set of candidate task-related text feature vectors $\bm{t}\in\mathbb{R}^{G \times N}$ at the receiver, where $G$ is the number of text vectors. Each text feature vector is extracted using the pre-trained CLIP text encoder in the semantic space from a human-readable task-relevant text prompt (e.g. “a photo of a cat” for image classification), enabling comparison with the decoded image vector $\bm{\hat{s}}$ via similarity computation. The process can be defined as
\begin{equation} \label{task} \bm{\hat{p}} = \text{Softmax}(\text{cos}(\bm{\hat{s}}, \bm{t}^{\text{T}})), \end{equation}
\textcolor{black}{where $\bm{\hat{p}}\in \mathbb{R}^{B \times G}$ represents the probability distribution over $G$ candidate text prompts for each of the $B$ visual inputs.} Here,  $\text{cos}(\cdot, \cdot)$ denotes the cosine similarity. Each row of $\bm{{t}}$ is generated from a prompt by the CLIP text encoder in a transmission-aware manner, as detailed later in Eq. \eqref{text}. \textcolor{black}{This unified similarity-based formulation supports all multi-modal alignment tasks, such as image classification by associating each text prompt with a category, and cross-modal retrieval by retrieving relevant images given a text query prompt based on visual-textual similarity scores.}

\vspace{-1.2em}
\color{black}
\subsection{SNR-adaptive JSCC}
\begin{figure*}[t]
  \centering
  \centerline{\includegraphics[width=14cm]{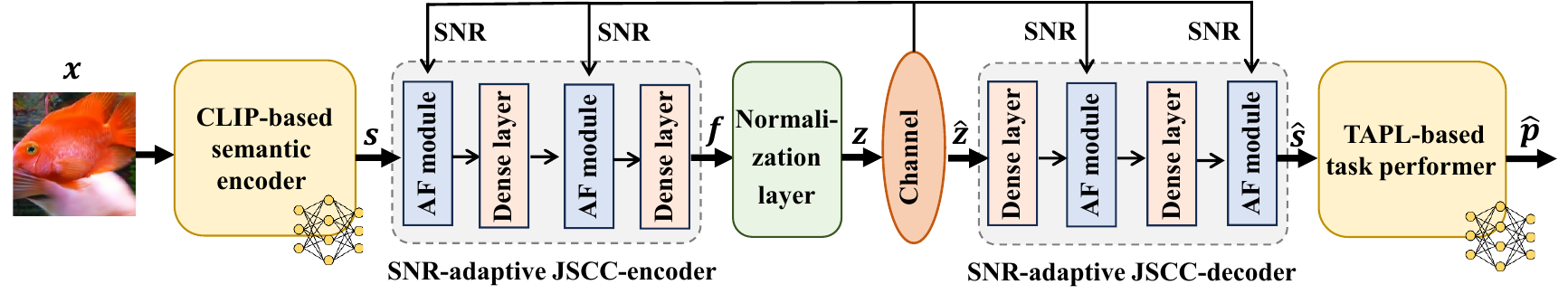}}
  \captionsetup{justification=raggedright}
  \caption{\textcolor{black}{Outline of the SemCLIP scheme: The image is encoded using a CLIP-based semantic encoder and transmitted via a DeepJSCC scheme. At the receiver, a TAPL mechanism is proposed to mitigate channel noise effects and enhance performance.}}
  \vspace{-1.5em}
  \label{SemCLIP framework}
\end{figure*}

An SNR-adaptive JSCC approach~\cite{CA2,xu2021wireless} is proposed to facilitate SNR-adaptive CLIP image token transmission. As shown in Fig.~\ref{SemCLIP framework}, the JSCC-encoder/decoder comprises multiple dense layers and attention feature (AF) modules, where the intermediate features from dense layers are concatenated with the SNR as input to the AF module, enabling feature scaling based on SNR. SNR is randomly sampled from a specified range during training, allowing the well-trained JSCC model to adapt to varying SNRs. Consequently, the distribution of the decoded image token vector at the receiver is inherently influenced by channel conditions and the JSCC encoding/decoding process during inference. \textcolor{black}{Unlike other DeepJSCC schemes, a contrastive learning loss is employed for robust transmission of task-agnostic semantic information, as detailed later.}

\vspace{-1.2em}
\subsection{Transmission-Aware Prompt Learning~(TAPL)}
\label{tapl}
Note that transmission over a noisy channel may alter the distribution of the original CLIP-based image token vector, potentially compromising the generalization of the reconstructed feature and preventing its perfect alignment with the corresponding text feature vector. 
To improve the efficiency of the multi-modal alignment of the task performer, a transmission-aware prompt learning method is proposed at the receiver. 

\begin{figure}
  \setlength{\abovecaptionskip}{0 cm}
  \setlength{\belowcaptionskip}{0cm}
  \centering
  \includegraphics[width=6cm]{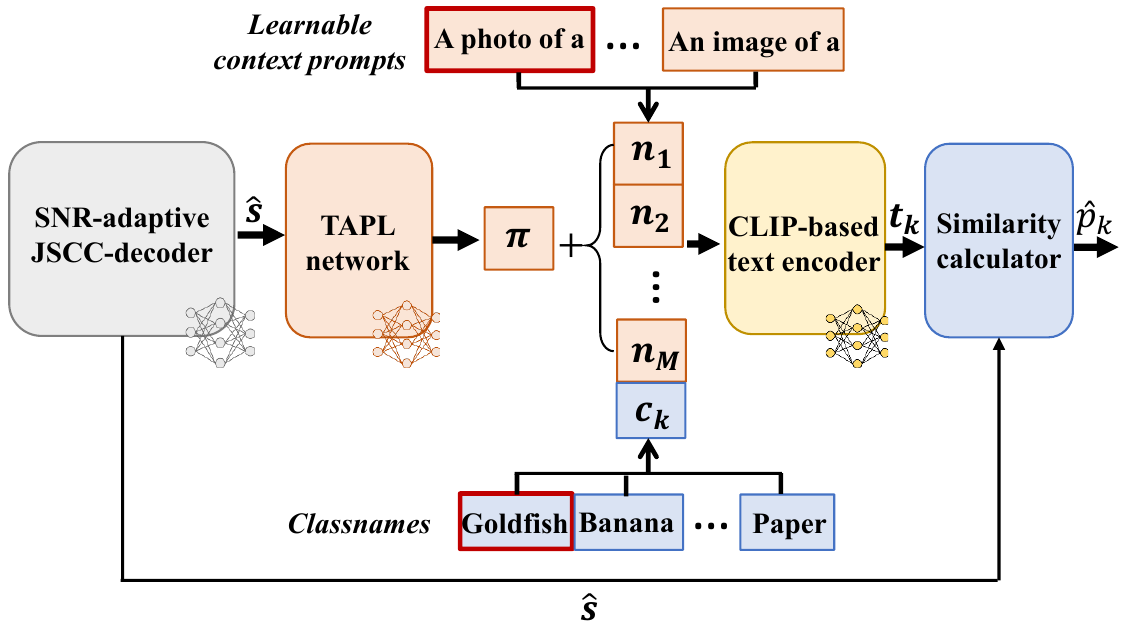}
  \captionsetup{justification=raggedright}
  \vspace{0pt}
  \caption{The structure of the TAPL-based task performer.}
  \vspace{-1.5em}
  \label{TAPL}
\end{figure}

Specifically, as shown in Fig.~\ref{TAPL}, the output of the JSCC-decoder $\bm{\hat{s}}$ is first fed into a TAPL network to be mapped to a conditional embedding vector $\bm{\pi}\in\mathbb{R}^{B \times D}$, where $D$ is the embedding dimension. $\bm{\pi}$ is conditioned on the decoded image vector $\bm{\hat{s}}$, which implicitly reflects the underlying channel conditions. The distribution of $\bm{\pi}$  can be used to adjust the text embeddings at the receiver according to channel conditions. These embeddings are obtained by mapping tokenized task-relevant prompts into an embedding space via the CLIP model’s embedding layer, and then fed into the text encoder to generate the final text feature vectors for semantic matching. The receiver has access to the full set of candidate task-related text prompts in advance. For example, the prompts can be ``a photo of a \{classname\}" for image classification, where ``a photo of a'' serves as the context prompt, and ``\{classname\}'' represents a candidate category to be classified. Then, to enable the text embeddings of prompts to better adapt to the reconstructed image tokens after transmission over the channel, thereby enhancing the effectiveness of downstream tasks, we model the embeddings of the context prompts using learnable continuous vectors, which are optimized through E2E training. On this basis, the distribution of received image vector is integrated into prompt learning by combining $\bm{\pi}$ with context embedding vectors. Consequently, the encoded text feature vector can be formulated as 
\begin{equation}
\label{text}
\bm{t}_k = T\bigl(Concat[(\bm{\pi}+ \bm{e}_1),(\bm{\pi}+ \bm{e}_2),...,(\bm{\pi}+ \bm{e}_M), \bm{c}_k]\bigl),
\end{equation}
where $[\bm{e}_1,...,\bm{e}_M]\in\mathbb{R}^D$ are the $M$ learnable context embedding vectors, $\bm{t}_k\in\mathbb{R}^N$ is the encoded feature vector of the $k$-th text prompt, and $\bm{c}_k\in\mathbb{R}^D$ represents the embedding vector of the $k$-th classname. 
Moreover, $Concat(\cdot)$ and $T(\cdot)$ denote the concatenation operation and the CLIP-based text encoding, respectively. Since $\bm{\pi}$ is obtained from the received $\bm{\hat{s}}$ decoded by the SNR-adaptive JSCC-decoder, it inherently captures the degradation and distortion induced by the channel. Therefore, the resulting text feature vectors derived from $\bm{\pi}$ become transmission-aware, adapting their semantic representations to better align with the image features under varying channel conditions.  \textcolor{black}{Task prediction probability is calculated as in Eq.~(\ref{task}).}

\vspace{-1.2em}
\subsection{Training Strategy}
\label{training strategy}
A two-stage training strategy is employed in this correspondence. The original image and text encoders of CLIP are used as the initial models and kept frozen during training~{\cite{clip}}. In the first stage, the SNR-adaptive JSCC encoder/decoder pair is trained while keeping other modules frozen to improve the reconstruction performance of image token transmission. Specifically, a contrastive loss {is adopted to learn a joint representation space in DeepJSCC for the transmitted and reconstructed image tokens~\cite{clip}:}
\begin{equation}
\label{deqn_ex1}
L_{\text{JSCC}}=\Bigl(\mathcal{L}_{ce}\bigl(\text{cos}(\bm{s},\hat{\bm{s}}^{\text{T}}),\bm{l}\bigl)+\mathcal{L}_{ce}\bigl(\bm{\text{cos}}(\hat{\bm{s}},\bm{s}^{\text{T}}),\bm{l}\bigl)\Bigl)/2,
\end{equation}
where $\text{cos}(\bm{s},\hat{\bm{s}}^{\text{T}})$ and $\text{cos}(\hat{\bm{s}},\bm{s}^{\text{T}})\in\mathbb{R}^{B \times B}$ are the cosine similarity matrices between the transmitted $\bm{s}$ and received $\hat{\bm{s}}$, respectively, $\bm{l} \triangleq [0,1, \cdots, B{-}1]
$ represents a label vector. The cross-entropy loss $\mathcal{L}_{ce}$ is defined as
$\mathcal{L}_{ce}\bigl(\text{cos}(\bm{s}, \hat{\bm{s}}^{\text{T}}), \bm{l} \bigr) = -\frac{1}{B} \sum_{i=1}^{B} \log \Bigl( \frac{e^{ \text{cos}(\bm{s}, \hat{\bm{s}}^{\text{T}})_{i, l_i}}}{\sum_{j=1}^{B} e^{\text{cos}(\bm{s}, \hat{\bm{s}}^{\text{T}})_{i, j}}} \Bigl)$,
where $l_i$ is the $i$-th element of $\bm{l}$, indicating the correct index for the $i$-th transmitted feature vector, $j$ indexes all reconstructed vectors as potential matching candidates. 
Similarity between image vectors of the transceiver corresponding to the same image is maximized, while minimized for pairs corresponding to different images.

To facilitate transmission-aware multi-modal alignment at the receiver, the TAPL network is trained with other modules frozen. Specifically, the contrastive loss for image-text pairs is utilized to align the two modalities~\cite{cocoop} as:
\begin{equation}
\label{deqn_ex1}
L_{\text{TAPL}}=\Bigl(\mathcal{L}_{ce}\bigl(\text{cos}(\hat{\bm{s}},\bm{t}^{\text{T}}),\bm{l}^{gt}\bigl)+\mathcal{L}_{ce}\bigl(\text{cos}(\hat{\bm{s}}^{\text{T}},\bm{t}),\bm{l}^{gt}\bigl)\Bigl)/2,
\end{equation}
where $\bm{l}^{gt}$ represents the ground-truth label vector, in which each element is an integer index indicating the correct target prompt for each input sample, corresponding to its position in the set of $G$ candidate prompts. $\bm{t}$ is generated from the learned prompt embeddings as in Eq.~(\ref{text}).

\begin{figure*}[t]
    \centering
    \begin{minipage}{0.43\textwidth}
    \centering
    \begin{minipage}{0.48\textwidth}
        \centering
        \includegraphics[width=\linewidth]{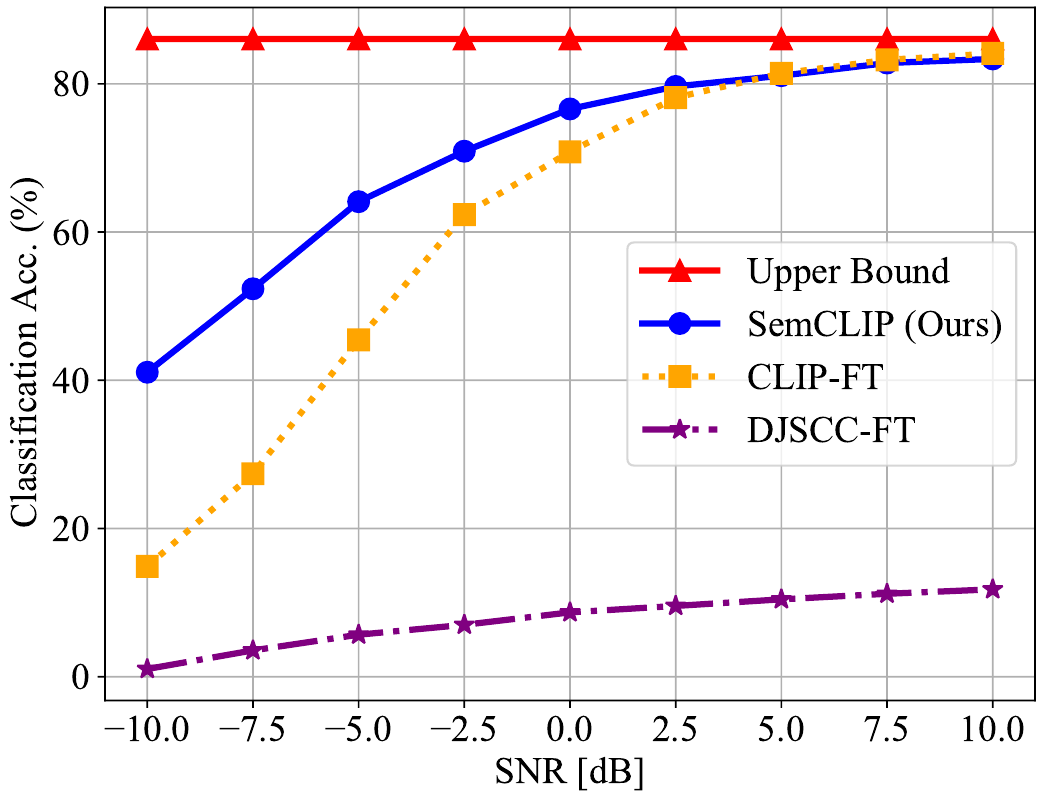}
        \caption*{(a) Classification.}
    \end{minipage}%
    \hfill
    \begin{minipage}{0.47\textwidth}
        \centering
        \includegraphics[width=\linewidth]{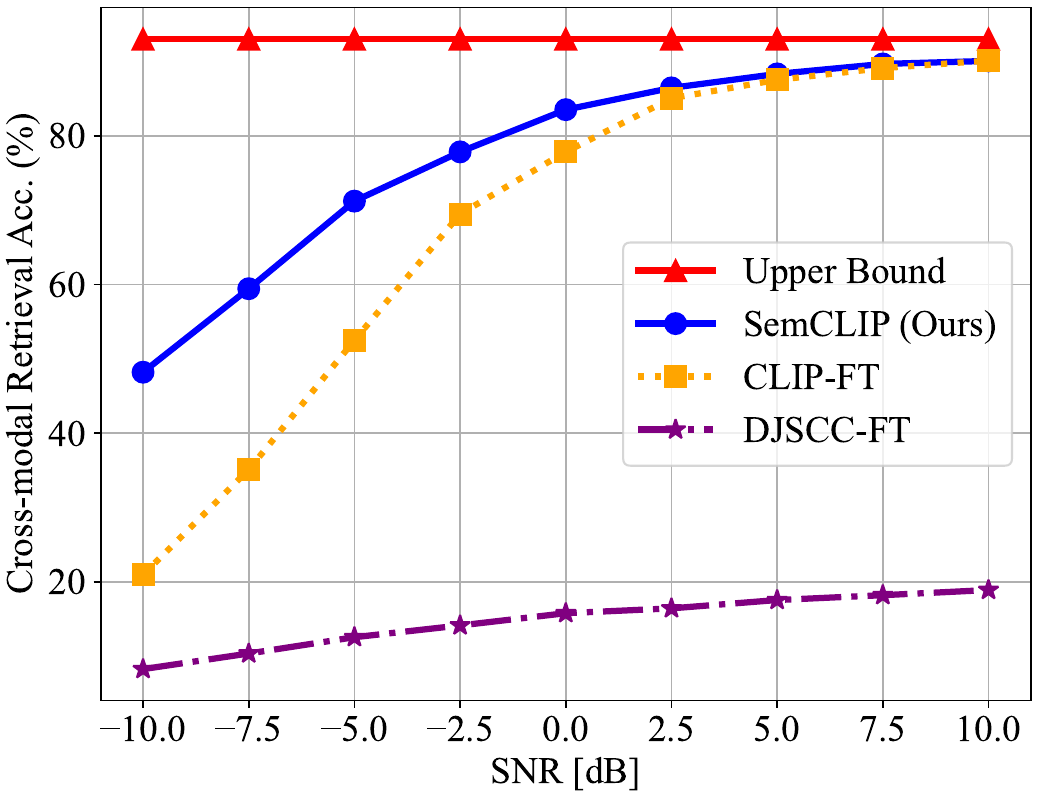}
        \caption*{(b) Cross-modal retrieval.}
    \end{minipage}
    \caption{\textcolor{black}{Zero-shot task performance vs. test channel SNR.}}
    \label{clip-jscc}
\end{minipage}
    \hspace{0.01\textwidth} 
    \begin{minipage}{0.225\textwidth} 
        \centering
        \includegraphics[width=\textwidth]{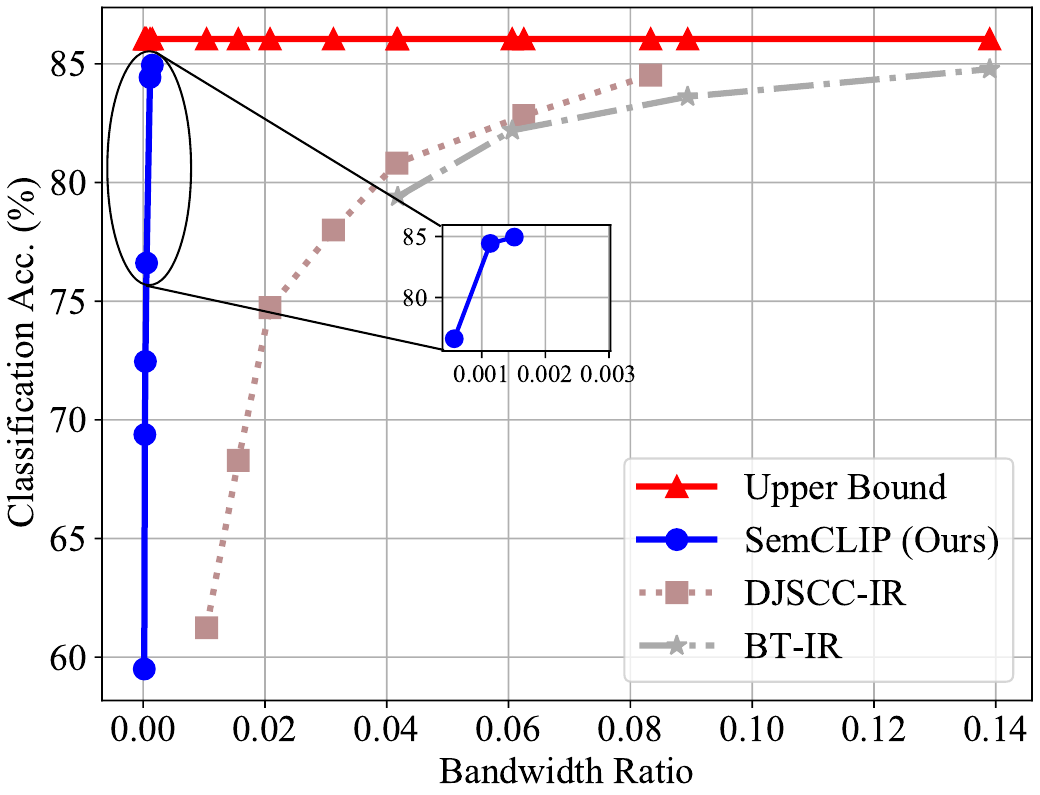}
        \caption{Zero-shot classification accuracy vs. bandwidth ratio.}
        \label{ratio}
    \end{minipage}
    \hspace{0.01\textwidth} 
    \begin{minipage}{0.3\textwidth} 
        \centering
    \includegraphics[width=\textwidth]{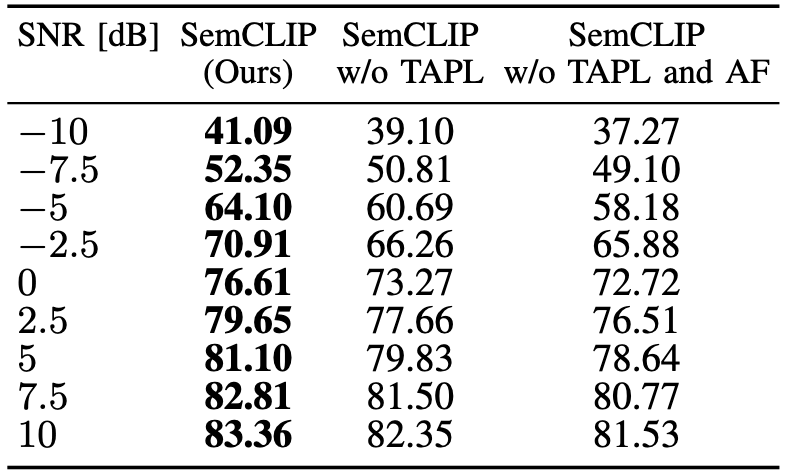}        
        \captionof{table}{Zero-shot classification comparison of different models vs. various SNRs.}
        \label{ablation}
    \end{minipage}
\end{figure*}

\vspace{-0.8em}
\section{Simulation Results}

\subsection{Experimental Setup}
We use the ImageNet-$1$k dataset for training. To evaluate the zero-shot performance of SemCLIP on unseen categories, we shuffle the categories in ImageNet and use $800$ categories for training and the remaining unseen $200$ categories for testing. Additionally, we evaluate its cross-dataset performance on OxfordPets, Food101, and Caltech101 datasets. 
 If not specified otherwise, we employ CLIP ViT-L/14@336px backbone, which resizes the input images to $336 \times 336$, with $N=768$, $B=128$ and $L=384$. The context prompt length is fixed at 4, with initialization from the pre-trained word embeddings of “a photo of a” for prompt learning. 
The proposed SemCLIP scheme is compared with the following benchmarks in terms of the zero-shot task performance. 
 \begin{itemize}
         \item[$\bullet$] {IR-based methods: 1) DJSCC-IR: images are transmitted using the state-of-the-art transformer-based DeepJSCC scheme \cite{wu2024transformer}. 2) Bit transmission~(BT)-IR: images are transmitted through a separation-based digital scheme, where the image is first compressed using the learned compression scheme from~\cite{cheng2020learned}, followed by capacity-achieving channel codes.
        The reconstructed images from these methods are fed into the CLIP encoder for zero-shot task at the receiver. }
         
        \item[$\bullet$] {FT-based methods: 1) CLIP-FT: the original CLIP-encoded tokens are assumed to be available to the receiver, serving as an upper bound on the performance of any CLIP-based scheme. 
        2) DJSCC-FT: image features are extracted and encoded by a deep neural network~(DNN)-based semantic encoder and trasmitted using a DeepJSCC encoder-decoder pair as in~\cite{lo2023collaborative,jscc1-retrival}, performing the task directly using the reconstructed features at the receiver. The same ImageNet category split as in SemCLIP is used for training and testing. } 
        
\end{itemize}

\begin{table}[t]
\caption{{Zero-shot classification accuracy for various datasets.}}
\centering
\tabcolsep=0.1cm
\renewcommand\arraystretch{0.65}
{
\fontsize{9.5}{11}\selectfont  
\begin{tabular}{lccc}
\toprule
\textnormal{\fontsize{9}{10}\selectfont Methods} & 
\textnormal{\fontsize{9}{10}\selectfont OxfordPets} & 
\textnormal{\fontsize{9}{10}\selectfont Food101} & 
\textnormal{\fontsize{9}{10}\selectfont Caltech101} \\
\midrule
Upper bound   & 90.99 & 91.12 & 88.83  \\
SemCLIP   & \textbf{82.28} & \textbf{82.87} & \textbf{80.47}  \\
SemCLIP w/o TAPL    & 79.37 & 78.75 & 77.95  \\
SemCLIP w/o TAPL and AF  & 77.51 & 77.47 & 77.43  \\
CLIP-FT   &77.00  & 77.38 & 77.39\\
\bottomrule
\end{tabular}}
\label{datasets}
\vspace{-0.6em}
\end{table}

\begin{table}[t]
\centering
\caption{{Computational complexity and execution delay for processing 100 images of each module in SemCLIP.}}
\renewcommand{\arraystretch}{1.1} 
\begin{tabular}{|>{\centering\arraybackslash}m{3.5cm}|>{\centering\arraybackslash}m{2cm}|>{\centering\arraybackslash}m{2cm}|}
\hline
\textbf{Module} & \textbf{Parameters} & \textbf{Execution delay} \\ \hline
CLIP-based semantic encoder & $304.3~\text{M}$ & $1.84~\text{s} $ \\ \hline
SNR-adaptive JSCC-encoder & 1.29~\text{M}  & $0.23~\text{ms}$ \\ \hline
SNR-adaptive JSCC-decoder & 1.35~\text{M}  & $0.25~\text{ms}$ \\ \hline
TAPL-based task performer & $123.07~\text{M}$   & $122.4~\text{ms}$ \\ \hline
\end{tabular}
\vspace{-1.5em}
\label{table:complexity}
\end{table}

\vspace{-1.5em}
\subsection{Numerical Experiments}

\subsubsection{Zero-shot experiments}
\textcolor{black}{Fig.~\ref{clip-jscc} presents the zero-shot image classification and cross-modal retrieval accuracy on unseen categories of different methods across various SNRs. Direct task performance based on the original CLIP vectors without transmission errors serves as an upper bound. It is obvious that the performance of SemCLIP is superior to other feature transmission-based methods, especially at low SNR regimes. Specifically, at $-5$ dB, SemCLIP achieves a $41\%$ performance gain compared to CLIP-FT on classification, in terms of cross-modal retrieval, the gap is $35.67\%$.} The image tokens from SemCLIP are better reconstructed from noisy transmission using SNR-adaptive JSCC, and the text features align more effectively with the image features in high-dimensional space using transmission-aware fine-tuned prompts. Note that the  DJSCC-FT fails to accurately classify unseen categories across different SNR levels, demonstrating that the existing DeepJSCC-based SemCom fails to handle task-agnostic scenarios. Moreover, at high SNR regime, the performance of SemCLIP nearly reaches the upper bound even though the transmitted vector is highly compressed, illustrating the effectiveness of the proposed CLIP-based JSCC and TAPL.

To verify the transmission efficiency of SemCLIP, zero-shot classification accuracy of different methods is presented in Fig.~\ref{ratio} as a function of the bandwidth ratio at SNR $= 0$ dB, where the bandwidth ratio is defined as $R \triangleq \frac{L}{C \times W \times H}$. The DJSCC-IR and BT-IR methods can only work at very high bandwidth ratios, while SemCLIP achieves close-to-optimal performance at extremely low bandwidth ratios. Specifically, the zero-shot accuracy of SemCLIP reaches $84.95\%$ when the bandwidth ratio is $0.0015$, whereas DJSCC-IR and BT-IR require bandwidth ratios of $0.0833$ and $0.139$, respectively, to achieve accuracies of $84.53\%$ and $84.78\%$. In other words, the proposed SemCLIP can save bandwidth nearly $55$ and $92$ times compared to DJSCC-IR and BT-IR, respectively, thanks to highly integrated and generalized  CLIP-based JSCC encoding. 

\subsubsection{Performance of TAPL} 
We perform an ablation study for comparison to assess the impact of the TAPL and AF modules. Table~\ref{ablation} presents the zero-shot classification accuracy on unseen categories as a function of the channel SNR. We observe that SemCLIP clearly achieves the best performance in all cases. Specifically, at $-10$ dB, the zero-shot performance has been improved by $5.09\%$ compared to SemCLIP without TAPL, which is attributed to efficient prompt learning. Meanwhile, SemCLIP without TAPL outperforms the CLIP-based JSCC method with specific SNR training by $4.91\%$, where the AF module and prompt learning are not considered. This demonstrates that SNR-adaptive JSCC improves both training efficiency, by training once across multiple SNR values, and CLIP feature encoding performance, by scaling intermediate vectors adaptively.

\subsubsection{Generalizability} Finally, the cross-dataset performance of different methods is studied in Table~\ref{datasets}, where the pre-trained models are tested on new datasets directly at SNR $= 0$ dB to further demonstrate the generalizability of SemCLIP. It shows that all the CLIP-based JSCC methods can be utilized on different datasets. Particularly, the zero-shot accuracy increases by $3.67\%, 6.15\%, 6.86\%$ on OxfordPets compared to SemCLIP w/o TAPL, SemCLIP w/o TAPL and AF, CLIP-FT, respectively. This is thanks to SemCLIP's capability to extract and convey universal features of images, rather than focusing on specific tasks, enhancing the generalizability and universality of SemCom systems.

\vspace{-1.5em}
\subsection{Computational and Implementation Complexity}
The number of parameters and execution delay for processing 100 images of different modules of SemCLIP are shown in Table~\ref{table:complexity}. The CLIP-based semantic encoder and TAPL-based task performer are the most computationally intensive modules, as they include transformer-based image and text encoders from the CLIP model with large parameter sizes. However, the parameters of CLIP encoders are frozen during training, and the CLIP image feature vectors are extracted in advance and stored  for the entire training. As a result, the CLIP image encoder is only executed once for each image, reducing the computational overhead significantly. Complexity of SemCLIP can be further reduced by using recently developed simplified CLIP models that achieve remarkable performance with significantly reduced number of parameters~\cite{liu2024simplifying}.

\vspace{-1.2em}
\section{Conclusions}
\textcolor{black}{In this correspondence, we proposed SemCLIP, a foundation model-driven SemCom system leveraging the generalizability of the CLIP model. Specifically, an SNR-adaptive DeepJSCC framework is designed to transmit generalized CLIP image tokens efficiently. To improve the accuracy of multi-modal alignment of the task performer at the receiver, a TAPL method is proposed to dynamically adjust text prompts based on the channel output distributions. Simulation results demonstrate that SemCLIP outperforms all alternative baselines by a large margin, achieving a $41\%$ improvement in zero-shot classification accuracy on unseen categories at very low SNRs, and can significantly reduce bandwidth requirements. The proposed scheme provides a unified approach easily extendable to token transmission for other foundation models, highlighting the potential of foundation models for a generalized SemCom over wireless networks.}

\footnotesize{
\bibliographystyle{IEEEtran}
\bibliography{IEEEabrv,clip}}
\newpage






\end{document}